\documentclass[12pt]{article}

\usepackage{graphicx}% Include figure files

\usepackage{times}

\newenvironment{sciabstract}{%
\begin{quote} \bf}
{\end{quote}}

\newcounter{lastnote}
\newenvironment{scilastnote}{%
\setcounter{lastnote}{\value{enumiv}}%
\addtocounter{lastnote}{+1}%
\begin{list}%
{\arabic{lastnote}.} {\setlength{\leftmargin}{.22in}}
{\setlength{\labelsep}{.5em}}} {\end{list}}

\title{Prospects for Strong Cavity Quantum Electrodynamics
 with Superconducting Circuits
}

\author
{S. M. Girvin$^1$, Ren-Shou Huang$^{1,2}$, Alexandre Blais$^1$,\\
Andreas Wallraff$^3$ and  R. J. Schoelkopf$^3$
\\
\normalsize{$^{1}$Department of Physics, Sloane Physics
Laboratory, Yale
University, New Haven, CT 06520-8120}\\
\normalsize{$^2$Department of Physics, Indiana University,
Bloomington, IN 47405}\\
 \normalsize{$^{3}$Department of Applied
Physics, Becton Laboratory, Yale University, New Haven, CT
06520-8284}\\
\\
}
\date{\today}

\def\be{\begin{equation}}
\def\ee{\end{equation}}
\def\omegar{\omega_{\rm r}}

\def\ndet{n_{\rm amp}}
\def\ncrit{n_{\rm crit}}

\begin{document}

\baselineskip24pt

\maketitle

%\begin{abstract}
\begin{sciabstract}
We propose a realizable architecture using one-dimensional
transmission line resonators to reach the strong coupling limit of
cavity quantum electrodynamics in superconducting electrical
circuits. The vacuum Rabi frequency for the coupling of cavity
photons to quantized excitations of an adjacent electrical circuit
(qubit) can easily exceed the damping rates of both the cavity and
the qubit.  This architecture is attractive for quantum computing
and control, since it provides strong inhibition of spontaneous
emission, potentially leading to greatly enhanced qubit lifetimes,
allows high-fidelity quantum non-demolition measurements of the
state of multiple qubits, and has a natural mechanism for
entanglement of qubits separated by centimeter distances. In
addition it would allow production of microwave photon states of
fundamental importance for quantum communication.
\end{sciabstract}
%\end{abstract}

\newpage

%
%%%%%%%%%%%%%%%%%%%%%%%%%%%
% I: Intro and cQED Review
%%%%%%%%%%%%%%%%%%%%%%%%%%%

\begin{figure}[ht]
\begin{center}
\includegraphics[width=4
in]{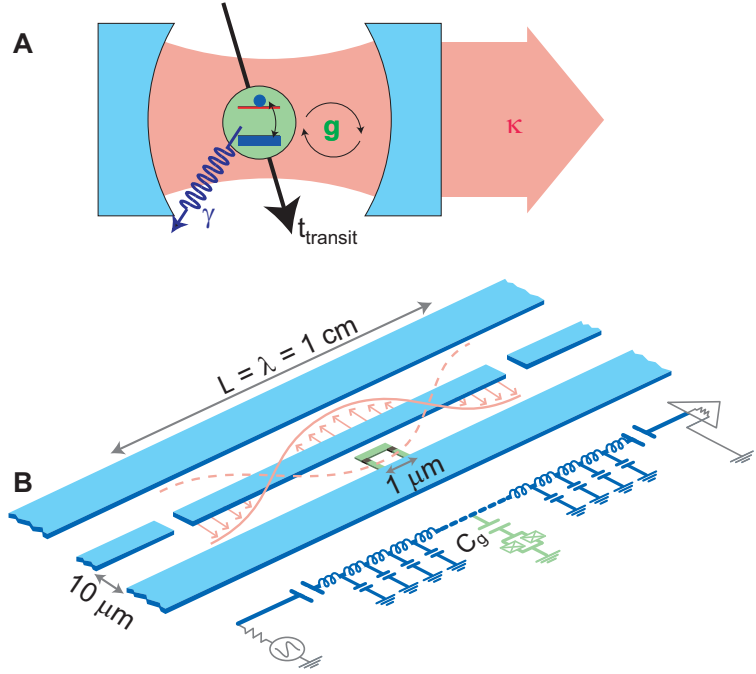} \caption{a) Standard representation of
cavity quantum electrodynamic system, comprising a single mode of
the electromagnetic field in a cavity with decay rate $\kappa$
coupled with a coupling strength $g={\cal E}_{\rm rms}d/\hbar$ to
a two-level system with spontaneous decay rate $\gamma$ and cavity
transit time $t_{\rm transit}$. b) Schematic layout and effective
circuit of proposed implementation of cavity QED using
superconducting circuits. The 1D transmission line resonator
consists of a full-wave section of superconducting coplanar
waveguide, which may be lithographically fabricated using
conventional optical lithography. A Cooper-pair box qubit is
placed between the superconducting lines, and is capacitively
coupled to the center trace at a maximum of the voltage standing
wave, yielding a strong electric dipole interaction between the
qubit and a single photon in the cavity. The box consists of two
small ($\sim 100 \, \rm{nm} \times 100 \, \rm{nm}$) Josephson
junctions, configured in a $\sim 1 \, \rm{\mu m}$ loop to permit
tuning of the effective Josephson energy by magnetic field. Input
and output signals are coupled to the resonator, via the
capacitive gaps in the center line, from $50 \, \Omega$
transmission lines which allow measurements of the amplitude and
phase of the cavity transmission, and the introduction of dc and
rf pulses to manipulate the qubit states. Multiple qubits (not
shown) can be similarly placed at different antinodes of the
standing wave to generate entanglement and two-bit quantum gates
across distances of several millimeters.}
\label{fig:resonatorandbox}
\end{center}
\end{figure}

Cavity quantum electrodynamics (cQED) studies the properties of
atoms coupled to discrete photon modes in high $Q$ cavities.  Such
systems are of great interest in the study of fundamental quantum
mechanics of open systems, the engineering of quantum states and
the study of measurement-induced decoherence
\cite{mabuchi:2002,hood:2002,raimond:2001}, and have also been
proposed as possible candidates for use in quantum information
processing and transmission
\cite{mabuchi:2002,hood:2002,raimond:2001}. Ideas for novel cQED
analogs using nano-mechanical resonators have recently been
suggested by Schwab and collaborators
\cite{armour:2002,elinor:2003}. We present a realistic proposal
for cQED via Cooper pair boxes coupled to a one-dimensional (1D)
transmission line resonator as shown in Fig.
\ref{fig:resonatorandbox}, within a simple circuit that can be
fabricated on a single microelectronic chip. As we discuss, 1D
cavities offer a number of practical advantages in reaching the
strong coupling limit of cQED over previous proposals using
discrete LC circuits \cite{makhlin:2001,buisson:2001}, large
Josephson junctions
\cite{marquandt:2001,plastina:2003,blais:2003}, or 3D cavities
\cite{al-saidi:2001,yang:2003,younoria:2003}.  Besides the
potential for entangling qubits to realize two-qubit gates
addressed in those works, we show that the cQED approach also
gives strong and controllable isolation of the qubits from the
electromagnetic environment, permits high fidelity quantum
non-demolition (QND) readout of multiple qubits, and can produce
states of microwave photon fields suitable for quantum
communication. The proposed circuits therefore provide a simple
and efficient architecture for solid-state quantum computation, in
addition to opening up a new avenue for the study of entanglement
and quantum measurement physics with macroscopic objects. We will
frame our discussion in a way that makes contact between the
language of atomic physics and that of electrical engineering, and
begin with a brief general overview of cQED before turning to a
more specific discussion of our proposed architecture.

In the optical version of cQED \cite{hood:2002}, one drives the
cavity with a laser and monitors changes in the cavity
transmission resulting from coupling to atoms falling through the
cavity.  One can also monitor the spontaneous emission of the
atoms into transverse modes not confined by the cavity.  It is not
generally possible to directly determine the state of the atoms
after they have passed through the cavity because the spontaneous
emission lifetime is on the scale of nanoseconds. One can,
however, infer information about the state of the atoms inside the
cavity from real-time monitoring of the cavity optical
transmission.

In the microwave version of cQED \cite{raimond:2001} one uses a
very high $Q$  superconducting 3D resonator to couple photons to
transitions in Rydberg atoms.  Here one does not directly monitor
the state of the photons, but is able to determine with high
efficiency the state of the atoms after they have passed through
the cavity (since the excited state lifetime is of order 30 ms).
{From} this state-selective detection one can infer information
about the state of the photons in the cavity.

The key parameters describing a cQED system (see Table I) are the
cavity resonance frequency $\omegar$, the atomic transition
frequency $\Omega$, and the strength of the atom-photon coupling
$g$ appearing in the Jaynes-Cummings \cite{walls-milburn}
Hamiltonian \be H = \hbar\omegar \left(a^\dagger a +
\frac{1}{2}\right) + \frac{\hbar\Omega}{2}\sigma^z + \hbar
g(a^\dagger\sigma^-+a\sigma^+) + H_\kappa+H_{\gamma}.
\label{eq:Jaynes-Cummings} \ee Here $H_\kappa$ describes the
coupling of the cavity to the continuum which produces the decay
rate $\kappa = \omegar/Q$, while $H_{\gamma}$ describes the
coupling of the atom to modes other than the cavity mode which
cause the excited state to decay at rate $\gamma$ (and possibly
also produce additional dephasing effects). An additional
important parameter in the atomic case is the transit time
$t_{\rm{transit}}$ of the atom through the cavity. In the absence
of damping and for the case of zero detuning
[$\Delta\equiv\Omega-\omegar=0$] between the atom and the cavity,
an initial zero-photon excited atom state
$\left|0,\uparrow\right\rangle$ flops into a photon
$\left|1,\downarrow\right\rangle$ and back again at the vacuum
Rabi frequency $g/\pi$. The degeneracy of the two corresponding
states with $n$ additional photons is split by $2 \hbar g
\sqrt{n+1}$. Equivalently, the atom's state and the photon number
are entangled.  The value of $g={\cal E}_{\rm rms}d/\hbar$ is
determined by the transition dipole moment $d$ and the rms
zero-point electric field of the cavity mode. Strong coupling is
achieved when $g\gg \kappa,\gamma$ \cite{haroche:92}.

%%%%%%%%%%%%%%%%%%%%%%%%%%
% II: Our sc QED proposal
%%%%%%%%%%%%%%%%%%%%%%%%%%

%%%%%
We now consider in more specific detail the cQED setup illustrated
in Fig.~\ref{fig:resonatorandbox}.  A number of possible
superconducting quantum circuits could function as the `atom'. For
definiteness we focus on the Cooper pair box
\cite{bouchiat:98,makhlin:2001,vion:2002,lehnert:2003}. Unlike the
usual cQED case, these artificial `atoms' remain at fixed
positions indefinitely and so do not suffer from the problem that
the coupling $g$ varies with position in the cavity. An additional
advantage is that the zero-point energy is distributed over a very
small effective volume ($\approx 10^{-5}$ cubic wavelengths) for
our choice of a quasi-one-dimensional transmission line `cavity.'
This leads to significant rms voltages $ {V}_{\rm rms}^0 \sim
\sqrt{\hbar\omega_r / c L}$ between the center conductor and the
adjacent ground plane at the anti\-nodal positions, where $L$ is
the resonator length and $c$ is the capacitance per unit length of
the transmission line. At a resonant frequency of $10 \, \rm{GHz}$
($h \nu / k_B \sim 0.5 \, \rm{K}$) and for a $10 \,\rm{\mu m}$ gap
between the center conductor and the adjacent ground plane,
${V}_{\rm rms} \sim 2 \, \rm{\mu V}$ corresponding to electric
fields ${\cal E}_{\rm rms} \sim 0.2 \, \rm{V/m}$, some $100$ times
larger than achieved in the 3D cavity described in
Ref.~\cite{raimond:2001}. Thus, this geometry might also be useful
for coupling to Rydberg atoms \cite{lukin:2003}.

In addition to the small effective volume, and the fact that the
on-chip realization of cQED shown in
Fig.~\ref{fig:resonatorandbox} can be fabricated with existing
lithographic techniques, a transmission-line resonator geometry
offers other practical advantages over LC circuits or large
Josephson junctions. The qubit can be placed within the cavity
formed by the transmission line to strongly suppress the
spontaneous emission, in contrast to an LC circuit, where
radiation and parasitic resonances may be induced in the wiring.
Since the resonant frequency of the transmission line is
determined primarily by a fixed geometry, its reproducibility and
immunity to 1/f noise should be superior to Josephson junction
resonators. Finally, transmission line resonances in coplanar
waveguides with $Q\sim 10^6$ have already been demonstrated
\cite{day:2003}, suggesting that the internal losses can be very
low.  The optimal choice of the resonator $Q$ in this approach is
strongly dependent on the presently unknown intrinsic decay rates
of superconducting qubits. Here we assume the conservative case of
an overcoupled resonator with a $Q\sim 10^4$, which is preferable
for the first experiments.

Our choice of `atom', the Cooper pair box
\cite{bouchiat:98,makhlin:2001} is a mesoscopic superconducting
grain with a significant charging energy.  The two lowest charge
states having $N_0$ and $N_0+1$ Cooper pairs are coherently mixed
by Josephson tunnelling between the box and a reservoir (in this
case the resonator ground plane) leading to the two-level
Hamiltonian \cite{makhlin:2001} \be
H_Q = E_{\rm el} \sigma^x - \frac{E_{\rm J}}{2}\sigma^z.  %note I have rotated about y axis
\ee Here, we have chosen the spinor basis such that the box Cooper
pair number operator is \cite{morethantwostates}
$\hat{N}-N_0=(1+\sigma^x)/2$. The electrostatic energy is given by
$4E_{\rm c}({C_{\rm g}V_{\rm g}}/{2e}-{1}/{2})$, where $C_{\rm g}$
is the coupling capacitance between the box and the resonator,
$E_{\rm c}\equiv{e^2}/{2C_\Sigma}$ is the charging energy
determined by the total box capacitance and $E_{\rm J}$ is the
Josephson energy. Dc gating of the box can be conveniently
achieved by applying a bias voltage to the center conductor of the
transmission line.
%The gate voltage $V_{\rm g}= V_{\rm g}^0+v$ has a
In addition to the dc part $V_{\rm g}^{\rm{dc}}$ the gate voltage
has a quantum part $v=V^0_{\rm rms} (a^\dagger + a)$ from which we
obtain \be g=\frac{E_{\rm J}}{\sqrt{E_{\rm J}^2+E_{\rm el}^2}}
\frac{e}{\hbar}\beta\sqrt{\frac{\hbar \omegar}{c L}}, \ee where
$\beta\equiv {C_{\rm g}}/{C_\Sigma}$. At the charge degeneracy
point $E_{\rm el}=0$ (where $n_{\rm g}=C_{\rm g} V_{\rm
g}^{\rm{dc}} /2e = 1/2$), the two levels are split only by the
Josephson energy and the `atom' is highly polarizable, having
transition dipole moment $d \equiv {\hbar g}/{{\cal E}_{\rm rms}}
\sim 2\times 10^4$ atomic units ($ea_0$), or more than an order of
magnitude larger than even a typical Rydberg atom
\cite{haroche:92}. An experimentally realistic \cite{lehnert:2003}
coupling $\beta\sim 0.1$ leads to a vacuum Rabi rate $g/\pi\sim
100$~MHz, which is three orders of magnitude larger than in
corresponding atomic microwave cQED experiments
\cite{raimond:2001}.

A comparison of the experimental parameters for implementations of
cavity QED with optical and microwave atomic systems, and for the
proposed implementation with superconducting circuits, is
presented in Table I. We assume a relatively low $Q=10^4$ and a
worst case estimate, consistent with the bound set by previous
experiments (discussed further below), for the intrinsic qubit
lifetime of $1/\gamma \geq 2\,\rm{\mu s}$. The standard figures of
merit~\cite{kimble:94} for strong coupling are the critical photon
number needed to saturate the atom on resonance $m_0
=\gamma^2/2g^2\le 1\times 10^{-6}$ and the minimum atom number
detectable by measurement of the cavity output $N_0
=2\gamma\kappa/g^2\le 6\times 10^{-5}$. These remarkably low
values are clearly very favorable, and show that superconducting
circuits could access the interesting regime of very strong
coupling.

%%%%%%%%%%%%%%%%%%
% III: Case of Zero Detuning
%%%%%%%%%%%%%%%%%%

For the case of zero detuning and weak coupling $g<\kappa$, the
radiative decay rate of the qubit into the transmission line
becomes strongly {\em enhanced} by a factor of $Q$ relative to the
rate in the absence of the cavity \cite{haroche:92} because of the
resonant enhancement of the density of states at the atomic
transition frequency. In electrical engineering language, the
$\sim 50 \, \Omega$ external transmission line impedance is
transformed on resonance to a high value which is better matched
to extract energy from the qubit. For strong coupling, the first
excited state becomes a doublet with line width $(\kappa
+\gamma)/2$ since the excitation is half atom and half photon
\cite{haroche:92}. As can be seen from Table I, the coupling is so
strong that, even for the low $Q=10^4$ we have assumed,
$2g/(\kappa + \gamma) \sim 100$ vacuum Rabi oscillations are
possible, and the frequency splitting ($g/\pi \sim 100 \,
\rm{MHz}$) will be readily resolvable in the transmission spectrum
of the resonator. This spectrum can be observed in the same manner
employed in optical atomic experiments, with a continuous wave
measurement at low drive, and will be of practical use to find the
dc gate voltage needed to tune the box into resonance with the
cavity. Of more fundamental importance than this simple avoided
level crossing however, is the fact that the Rabi splitting scales
with the square root of the photon number, making the level
spacing anharmonic.  This should cause a number of novel
non-linear effects \cite{walls-milburn} to appear in the spectrum
at higher drive powers when the average photon number in the
cavity is large ($\langle n \rangle > 1$). A conservative estimate
of the noise energy for a 10 GHz cryogenic high electron mobility
(HEMT) amplifier is $n_{\rm amp}=k_BT_N/\hbar\omega=100$ photons,
so these spectral features should be readily observable in a
measurement time $t_{\rm meas}=n_{\rm amp}/\langle n
\rangle\kappa$, or only $\sim 16 \, \rm{\mu s}$ for $\langle n
\rangle\sim 1$.

%%%%%%%%%%%%%%%%%%%%%%%%%%%%%
% IV:  Case of Large Detuning I
%%%%%%%%%%%%%%%%%%%%%%%%%%%%%

For the case of strong detuning, the coupling to the continuum is
substantially reduced.  One can view the effect of the detuned
resonator as filtering out the vacuum noise at the qubit
transition frequency or, in electrical engineering terms, as
providing an impedance transformation which strongly {\em reduces}
the real part of the environmental impedance seen by the qubit.
For large detuning the qubit excitation spends only a small
fraction of its time as a photon \cite{haroche:92} so that  the
decay rate into the transmission line is only $\gamma_\kappa =
\left({g}/{\Delta}\right)^2\kappa\sim 1/(64 \, \rm{\mu s})$, much
less than $\kappa$.

One of the important motivations for this cQED experiment is to
determine the various contributions to the qubit decay rate
$\gamma$ so that we can understand their fundamental physical
origins as well as engineer improvements. Besides $\gamma_\kappa$,
there are two additional contributions to
$\gamma=\gamma_\kappa+\gamma_\perp+\gamma_{\rm NR}$. Here
$\gamma_\perp$ is the decay rate into photon modes other than the
cavity mode, and $\gamma_{\rm NR}$ is the rate of other (possibly
non-radiative) decays. Optical cavities are relatively open and
$\gamma_\perp$ is significant, but for 1D microwave cavities,
$\gamma_\perp$ is expected to be negligible (despite the very
large transition dipole). For Rydberg atoms the two qubit states
are both highly excited levels and $\gamma_{\rm NR}$ represents
(radiative) decay out of the two-level subspace. For Cooper pair
boxes, $\gamma_{\rm NR}$ is completely unknown at the present
time, but could have contributions from phonons, two-level systems
in insulating \cite{Simmonds:2003} barriers and substrates, or
thermally excited quasiparticles.

For Cooper box qubits {\em not} inside a cavity, recent
experiments \cite{lehnert:2003} have determined a relaxation time
$1/\gamma=T_1\sim 1.3 \, \rm{\mu s}$ despite the back action of
continuous measurement by a SET electrometer.  Vion et al.
\cite{vion:2002} found $T_1\sim 1.84 \, \rm{\mu s}$ (without
measurement back action) for their charge-phase qubit. The rate of
relaxation expected from purely vacuum noise (spontaneous
emission) is \cite{lehnert:2003,makhlin:2001} \be \gamma_\kappa =
\frac{E^2_{\rm J}}{{E_{\rm J}^2+E_{\rm el}^2}}
\left(\frac{e}{\hbar}\right)^2\beta^2 2\hbar\Omega \,{\rm
Re}[Z(\Omega)]. \ee It is difficult in most experiments to
precisely determine the real part of the high frequency
environmental impedance $Z(\Omega$) presented by the leads
connected to the qubit, but reasonable estimates
\cite{lehnert:2003} yield values of $T_1$ in the range of $1 \,
\rm{\mu s}$. Thus in these experiments, if there are non-radiative
decay channels, they are at most comparable to the vacuum
radiative decay rate (and may well be much less). Experiments with
a cavity will present the qubit with a simple and well controlled
electromagnetic environment, in which the radiative lifetime can
be enhanced with detuning to $1/\gamma_\kappa > 64 \, \rm{\mu s}$,
allowing $\gamma_{\rm NR}$ to dominate and yielding valuable
information about any non-radiative processes.

%%%%%%%%%%%%%%%%%%%%%%%%%%%%%
%  V: Case of Large Detuning II
%%%%%%%%%%%%%%%%%%%%%%%%%%%%%

For large detuning, making the unitary transformation
$U=\exp{\left[(g/\Delta)(a\sigma^+ - a^\dagger\sigma^-)\right]}$
and expanding to second order in $g$, approximately diagonalizes
the Hamiltonian (neglecting damping for the moment) \be
UHU^\dagger \approx \hbar\left[\omegar +
\frac{g^2}{\Delta}\sigma^z\right] a^\dagger a +
\frac{1}{2}\hbar\left[\Omega+\frac{g^2}{\Delta}\right]\sigma^z.
\label{eq:Jaynes-Cummings-diagonal} \ee We see that there is a
dispersive shift of the cavity transition by $\sigma_z
g^2/\Delta$, that is the qubit pulls the cavity frequency by $\pm
g^2/\kappa\Delta=\pm 2.5$ line widths for a 10\% detuning.  Exact
diagonalization \cite{haroche:92} shows that the pull becomes
power dependent and decreases in magnitude for cavity photon
numbers on the scale $n=n_{\rm crit}\equiv \Delta^2/4g^2\sim 100$.
In the regime of non-linear response, single-atom optical
bistability \cite{walls-milburn} can be expected when the drive
frequency is off resonance at low power but on resonance at high
power \cite{smgunpublished}.

The state-dependent pull of the cavity frequency by the qubit can
be used to entangle the state of the qubit with that of the
photons passing through the resonator. For $g^2/\kappa\Delta > 1$
the pull is greater than the line width and the microwave
frequency can be chosen so that the transmission of the cavity is
close to unity for one state of the qubit and close to zero for
the other \cite{note_pull}. For ${g^2}/{\kappa\Delta} \ll 1$ the
state of the qubit is encoded in the phase of the transmitted
microwaves. An initial qubit state $\left|\chi\right\rangle =
\alpha\left|\uparrow\right\rangle+\beta
\left|\downarrow\right\rangle$ evolves under microwave
illumination into the entangled state
$\left|\psi\right\rangle=\alpha \left|\uparrow,\theta\right\rangle
+ \beta\left|\downarrow,-\theta\right\rangle$, where $\tan\theta =
{2g^2}/{\kappa\Delta}$, and $\left|\pm\theta\right\rangle$ are
(interaction representation) coherent states with the appropriate
mean photon number and opposite phases. Such an entangled state
can be used to couple qubits in distant resonators and allow
quantum communication \cite{vanEnk:98}. If an independent
measurement of the qubit state can be made, then such states can
be turned into photon Schr\"odinger cats \cite{haroche:92}.

\begin{figure}[ht]
\begin{center}
\vspace*{-1.5cm}
\includegraphics[width=2.5in]{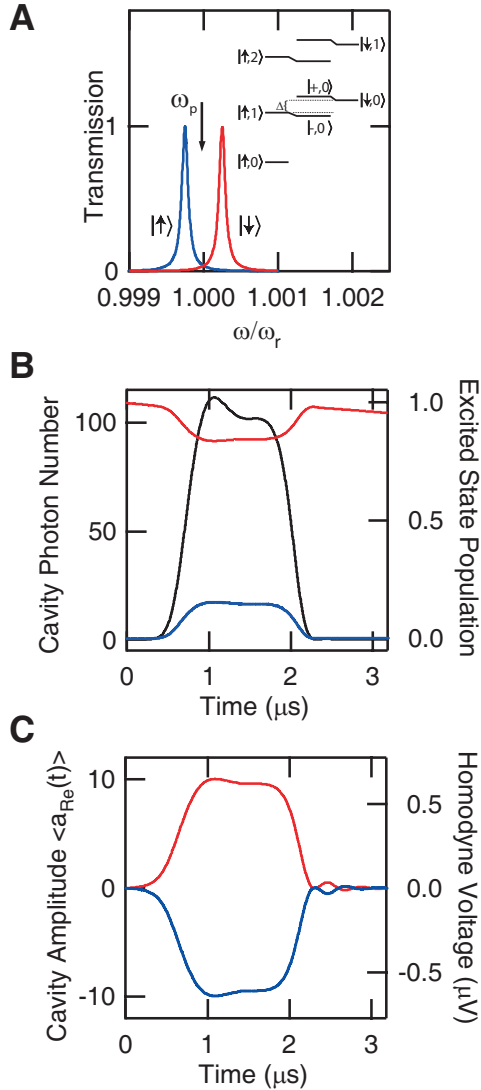}
\caption{\small{Use of the coupling between a Cooper-pair box
qubit and a transmission-line resonator to perform a dispersive
quantum non-demolition measurement. a) Transmission spectrum of
the cavity, which is "pulled" by an amount $\pm g^2/\Delta = 2.5
\times 10^{-4}\times \omega_r$, depending on the state of the
qubit (red for the excited state, blue for the ground state).  To
perform a measurement of the qubit, a pulse of microwave photons,
at a probe frequency $\omega_p=\omega_r$, is sent through the
cavity. Inset shows the dressed-state picture of energy levels for
the cavity-qubit system, for $10\,\%$ detuning. b) Results of
numerical simulations of this QND readout using the quantum state
diffusion method. A microwave pulse with duration $\sim 1.5 \,
\rm{\mu s}$ excites the cavity to an amplitude $\langle
n\rangle\sim 100$. The intracavity photon number (left axis, in
black), and occupation probability of the excited state, for the
case in which the qubit is initially in the ground (blue) or
excited (red) state, are shown as a function of time. Though the
qubit states are coherently mixed during the pulse, the
probability of real transitions is seen to be small. Depending on
the qubit's state, the pulse is either above or below the combined
cavity-qubit resonance, and so is transmitted with an large
relative phase shift that can be detected with homodyne detection.
c) The real component of the cavity electric field amplitude (left
axis), and the transmitted voltage phasor (right axis) in the
output transmission line, for the two possible qubit states. The
opposing phase shifts cause a change in sign of the output, which
can be measured with high signal-to-noise to realize a
single-shot, QND measurement of the qubit.}}
\label{fig:stochastic}
\end{center}
\end{figure}

The phase shift of the transmitted microwaves can be measured
using standard heterodyne techniques, and can therefore serve as a
high efficiency quantum non-demolition dispersive readout of the
state of the qubit, as described in Figure~\ref{fig:stochastic}.
Exciting the cavity to a maximal amplitude $\ncrit=100\sim \ndet$
the signal-to-noise ratio, SNR = $(\ncrit/\ndet)(\kappa/\gamma)$,
can be very high if the qubit lifetime is longer than a few cavity
decay times ($1/\kappa = 160 \,\rm{ns}$). We see from
Eq.~(\ref{eq:Jaynes-Cummings-diagonal}) that the ac-Stark/Lamb
shift of the box transition is $({2g^2}/{\Delta})(n+1/2)$, so the
back action of the dispersive cQED measurement is due to quantum
fluctuations of the number of photons in the cavity which cause
variations in the ac Stark shift, that dephase the qubit. A second
possible form of back action is mixing transitions between the two
qubit states induced by the microwaves. Since the coupling is so
strong, large detuning $\Delta = 0.1 \, \omegar$ can be chosen,
making the mixing rate limited not by the frequency spread of the
drive pulse, but rather by the width of the qubit excited state
itself. The rate of driving the qubit from ground to excited state
when $n$ photons are in the cavity is $R\approx
n(g/\Delta)^2\gamma$. If the measurement pulse excites the cavity
to $n=\ncrit$, we see that the excitation rate is still only 1/4
of the relaxation rate, so the main limitation on the fidelity of
the QND readout is the decay of the excited state of the qubit
during the course of the readout. This occurs (for small $\gamma$)
with probability $P_{\rm relax}\sim \gamma t_{\rm meas} \sim 5
\times \gamma/\kappa \sim 1.5 \, \%$ and the measurement is highly
non-demolition. The numerical stochastic wave function
calculations \cite{schack:97} shown in Fig.~\ref{fig:stochastic}
confirm that the measurement-induced mixing is negligible and that
one can determine the qubit's state in a single-shot measurement
with high fidelity. The readout fidelity, including the effects of
this stochastic decay, and related figures of merit of the QND
readout are summarized in Table II. Since nearly all the energy
used in this dispersive measurement scheme is dissipated in the
remote terminations of the input and output transmission lines, it
has the practical advantage of avoiding quasiparticle generation
in the qubit.

Another key feature of the cavity QED readout is that it lends
itself naturally to operation of the box at the charge degeneracy
point ($n_g=1/2$), where it has been shown that $T_2$ can be
enormously enhanced \cite{vion:2002} because the energy splitting
has an extremum with respect to gate voltage and isolation of the
qubit from 1/f dephasing is optimal. The derivative of the energy
splitting with respect to gate voltage is the charge difference in
the two qubit states.  At the degeneracy point this derivative
vanishes and the environment cannot distinguish the two states and
thus cannot dephase the qubit. This also implies that a charge
measurement cannot be used to determine the state of the system
\cite{armour:2002,elinor:2003}. While the first derivative of the
energy splitting with respect to gate voltage vanishes at the
degeneracy point, the second derivative, corresponding to the
difference in charge {\em polarizability} of the two quantum
states, is {\em maximal}.  One can think of the qubit as a
non-linear quantum system having a state-dependent capacitance (or
in general, an admittance) which changes sign between the ground
and excited states \cite{averin:2003}. % SMG REMOVED,averin:1985}.
It is this change in polarizability which is measured in the
dispersive QND measurement.

In contrast, standard charge measurement schemes
\cite{nakamura:99,lehnert:2003} require moving away from the
optimal point. Simmonds et al. \cite{Simmonds:2003} have recently
raised the possibility that there are numerous parasitic
environmental resonances which can relax the qubit when its
frequency $\Omega$ is changed during the course of moving the
operating point.  The dispersive cQED measurement is therefore
highly advantageous since it operates best at the charge
degeneracy point.  In general, such a measurement of an ac
property of the qubit is strongly desirable in the usual case
where dephasing is dominated by low frequency (1/f) noise. Notice
also that the proposed quantum non-demolition measurement would be
the inverse of the atomic microwave cQED measurement in which the
state of the photon field is inferred non-destructively from the
phase shift in the state of atoms sent through the cavity
\cite{raimond:2001}.

%%%%%%%%%%%%%%%%%%%%%%%%%%%%%%%%%%%%%%%%%%%%%%%%%%%%%%%%%%%%%%%%%
%  VI:  Multiple qubits
%%%%%%%%%%%%%%%%%%%%%%%%%%%%%%%%%%%%%%%%%%%%%%%%%%%%%%%%%%%%%%%%%

Finally, the transmission-line resonator has the advantage that it
should be possible to place multiple qubits along its length
($\sim 1 \, \rm{cm}$) and entangle them together, which is an
essential requirement for quantum computation. For the case of two
qubits, they can be placed closer to the ends of the resonator but
still well isolated from the environment and can be separately dc
biased by capacitive coupling to the left and right center
conductors of the transmission line. Any additional qubits would
have to have separate gate bias lines installed.  If qubits $i$
and $j$ are tuned in resonance with each other but detuned from
the cavity, the effective Hamiltonian will contain qubit-qubit
coupling due to exchange of virtual photons:
%\be
$H_2 = ({g^2}/{\Delta})(\sigma_i^+\sigma_j^- +
\sigma_i^-\sigma_j^+)$.
%\ee
Starting with an excitation in one of the qubits, this interaction
will have the pair of qubits maximally entangled after a time
$t_{\sqrt{i\rm SWAP}} = \pi\Delta/4g^2\sim 50 \, \rm{ns}$. Making
the most optimistic assumption that we can take full advantage of
the lifetime enhancement inside the cavity (i.e.~that $\gamma_{\rm
NR}$ can be made negligible), the number of $\sqrt{i\rm SWAP}$
operations which can be carried out in one cavity decay time is
$N_{\rm op} = 4\Delta/\pi\kappa\sim 1200$ for the experimental
parameters assumed above. This can be further improved if the
qubit's non-radiative decay is sufficiently small, and higher $Q$
cavities are employed. When the qubits are detuned from each
other, the qubit-qubit interaction in the effective Hamiltonian is
turned off, hence the coupling is tunable.
%SMG SINGLE BIT OPS
Numerical simulations indicate that when the qubits are strongly
detuned from the cavity, single-bit gate operations can be
performed with high fidelity \cite{smgunpublished}.
%SMG: CAN CUT THE PART BELOW HERE IF YOU WANT
Driving the cavity at its resonance frequency constitutes a {\em
measurement} because the phase shift of the transmitted wave is
strongly dependent on the state of the qubit and hence the photons
become entangled with the qubit. On the other hand, driving the
cavity at the qubit transition frequency constitutes a {\em
rotation}.  This is {\em not} a measurement because, for large
detuning the photons are largely reflected with a phase shift
which is independent of the state of the qubit. Hence there is
little entanglement and the rotation fidelity is high
\cite{smgunpublished}.

Together with one-qubit gates, the interaction $H_2$ is sufficient
for universal quantum computation (UQC) \cite{barenco:95}.
Alternatively, $H_2$ can be used to realize encoded UQC on the
subspace
$\mathcal{L}=\{|\!\!\uparrow\downarrow\rangle,|\!\!\downarrow\uparrow\rangle\}$
\cite{lidar:2002c}. In this context, a simpler non-trivial encoded
two-qubit gate can also be obtained by tuning, for a time  $t =
\pi\Delta/3g^2$, all four qubits in the pair of encoded logical
qubits in resonance with each other but detuned from the
resonator. This is closely related to the S\o rensen-M\o lmer
scheme discussed in the context of the ion-trap proposals
\cite{sorensen:99}.  Interestingly, $\mathcal{L}$ is also a
decoherence-free subspace with respect to global dephasing
\cite{lidar:2002c} and use of this encoding will provide some
protection against noise.

Another advantage of the dispersive QND readout is that one may be
able to determine the state of multiple qubits in a single shot
without the need for additional signal ports. For example, for the
case of two qubits with different detunings, the cavity pull will
take on four different values $\pm g_1^2/\Delta_1 \pm
g_2^2/\Delta_2$ allowing single-shot readout of the coupled
system.  This can in principle be extended to $N$ qubits provided
that the range of individual cavity pulls can be made large enough
to distinguish all the combinations. Alternatively, one could read
them out in small groups at the expense of having to electrically
vary the detuning of each group to bring them into strong coupling
with the resonator.

%%%%%%%%%%%%%%%%%%%%%%%%
% VII: Summary and Conclusions
%%%%%%%%%%%%%%%%%%%%%%%%

In summary, we propose that the combination of one-dimensional
superconducting transmission line resonators, which confine their
zero point energy to extremely small volumes, and superconducting
charge qubits, which are electrically controllable qubits with
large electric dipole moments, constitutes an interesting system
to access the strong-coupling regime of cavity quantum
electrodynamics. This combined system constitutes an advantageous
architecture for the coherent control, entanglement, and readout
of quantum bits for quantum computation and communication. Among
the practical benefits of this approach are the ability to
suppress radiative decay of the qubit while still allowing one-bit
operations, a simple and minimally disruptive method for readout
of single and multiple qubits, and the ability to generate tunable
two-qubit entanglement over centimeter-scale distances. We also
note that in the structures described here, the emission or
absorption of a single photon by the qubit is tagged by a sudden
large change in the resonator transmission properties
\cite{smgunpublished} making them potentially useful as single
photon sources and detectors.

\bibliographystyle{Science}

\begin{thebibliography}{10}
\bibitem{morethantwostates}
For a large ratio of $E_{\rm J}/E_{\rm c}$ higher charge states
will be mixed
  into the two lowest energy eigenstates and the coefficient of $\sigma_x$ in
  this expression will be somewhat different from unity.
\bibitem{note_pull}
We note that for the case of $Q=10^6$, the cavity pull is a
remarkable $\pm250$
  line widths, but, depending on the non-radiative decay rate of the qubit,
  this may be in the regime $\kappa < \gamma$, making the state measurement too
  slow.
\end{thebibliography}

\begin{scilastnote}
\item   We are grateful to David DeMille and Michel Devoret for
numerous conversations. This work was supported in part by the
National Security Agency (NSA) and Advanced Research and
Development Activity (ARDA) under Army Research Office (ARO)
contract number DAAD19-02-1-0045, NSF DMR-0196503, the David and
Lucile Packard Foundation, the W.M. Keck Foundation, NSERC and the
Canadian Foundation for Innovation.
\end{scilastnote}

%%%Table 1%%%%%%%%%%%%%%%%%%%%%%%%%%%%%%%%%%%%%%%%%%%%%%%%%%%%%%%%%%%%%%%%%%%%%%%%%%%%%%%%%%%%%%%%%%%%%%

\begin{table}[ht]

%\begin{center}
\begin{small}
\hspace*{-1.5cm}\begin{tabular}{|l|l|c|c|c|}\hline
parameter&symbol&3D optical&3D microwave&1D circuit\\ \hline %
resonance/transition frequency&$\omega_{\rm{r}}/ 2\pi$, $\Omega/ 2 \pi$&$350 \, \rm{THz}$&$51 \, \rm{GHz}$& $10 \, \rm{GHz}$\\ \hline %
vacuum Rabi frequency&$g/\pi$, $g/\omegar$&$220 \, \rm{MHz}$, $3 \times 10^{-7}$&$47 \, \rm{kHz}$, $1 \times 10^{-7}$&$100 \, \rm{MHz}$, $5\times 10^{-3}$\\ \hline %
transition dipole&$d/e a_0$&$\sim 1$&$1 \times 10^3$&$ 2\times 10^4$\\ \hline%
cavity lifetime&$1/\kappa, Q$&$10 \, \rm{ns}$, $3 \times 10^7$ & $1 \, \rm{ms}$, $3 \times10^{8}$&$160 \, \rm{ns}$, $10^4$\\ \hline %
atom lifetime &$1/\gamma$&$ 61\,\rm{ns}$&$30 \, \rm{ms}$&$ 2 \, \rm{\mu s}$\\ \hline %
atom transit time&$t_{\rm transit}$&$\ge 50 \, \rm{\mu s}$&$100 \, \rm{\mu s}$&$\infty$\\ \hline %
critical atom number&$N_0=2\gamma\kappa/g^2$&$6 \times 10^{-3}$&$3\times10^{-6}$&$\leq 6 \times 10^{-5}$\\ \hline %
critical photon number&$m_0=\gamma^2/2g^2$&$3\times10^{-4}$&$3\times10^{-8}$&$\leq 1 \times 10^{-6}$\\ \hline %
\# of vacuum Rabi flops&$n_{\rm Rabi}= 2g/(\kappa+\gamma)$ &$\sim 10$& $\sim 5$ &$\sim 10^2$ \\ \hline %
\end{tabular}
\end{small}
%\end{center}
\caption{Comparison of key rates and cQED parameters for optical
\cite{hood:2002} and microwave \cite{raimond:2001} atomic systems
using 3D cavities, compared against the proposed approach using
superconducting circuits, showing the possibility for attaining
the strong cavity QED limit ($n_{\rm Rabi} \gg 1$). For the 1D
superconducting system, a full-wave ($L=\lambda$) resonator,
$\omegar/2\pi=10$ GHz, a relatively low $Q$ of $10^4$ and coupling
$\beta=C_g/C_\Sigma=0.1$ are assumed. For the 3D microwave case,
the number of Rabi flops is limited by the transit time. For the
1D circuit case, the intrinsic Cooper-pair box decay rate is
unknown; a conservative value equal to the current experimental
upper bound $1/\gamma \ge 2 \, \rm{\mu s}$ is assumed. }
\end{table}
%%%%%%%%%%%%%%%%%%%%%%%%%%%%%%%%%%%%%%%%%%%%%%%%%%%%%%%%%%%%%%%%%%%%%%%%%%%%%%%%%%%%%%%%%%%%%

%%%%Table 2%%%%%%%%%%%%%%%%%%%%%%%%%%%%%%%%%%%%%%%%%%%%%%%%%%%%%%%%%%%%%%%%%%%%%%%%%%%%%%%%%%%%%%%%%%%%%

\begin{table}[ht]
\begin{center}
\begin{small}
\begin{tabular}{|l|c|c|}\hline

parameter&symbol&1D circuit\\ \hline %
dimensionless cavity pull&$g^2/\kappa\Delta$&2.5\\ \hline %
cavity-enhanced lifetime&$\gamma^{-1}_\kappa=(\Delta/g)^2\kappa^{-1}$&$64\ \mu$s \\ \hline
%%readout excitation&$P_{{\rm mix}\uparrow}=n_{\rm det}(g/\Delta)^2\gamma/\kappa$&$0.02; 6 \times 10^{-4}$\\ \hline %
%readout de-excitaton&$P_{{\rm mix}\downarrow}=\gamma/\kappa$&$0.08;\, 2.5\times 10^{-3}$\\ \hline %
readout SNR&SNR = $(\ncrit/\ndet)\kappa/\gamma$ &400 (12.5) \\ \hline %
readout error&$P_{\rm relax} \sim 5 \times \gamma/\kappa$ & 1.5 \% (14 \%)\\ \hline %
1 bit operation time& $T_\pi> 1/\Delta$ & $> 0.16 \, \rm{ns}$\\ \hline %
entanglement time&$t_{\sqrt{i{\rm SWAP}}} =\pi\Delta/4g^2$&$\sim 0.05\ \mu$s\\ \hline %
2 bit operations&$N_{\rm op}= 1/[\gamma\, t_{\sqrt{i{\rm SWAP}}}]$&$> 1200$ (40)\\ \hline %
\end{tabular}
\end{small}
\end{center}
\caption{Figures of merit for readout and multi-qubit entanglement
of superconducting qubits using dispersive (off-resonant) coupling
to a 1D transmission line resonator. The same parameters as Table
1, and a detuning of the Cooper pair box from the resonator of
10\% ($\Delta = 0.1 \, \omega_r$), are assumed. Quantities
involving the qubit decay $\gamma$ are computed both for the
theoretical lower bound $\gamma = \gamma_\kappa$ for spontaneous
emission via the cavity, and (in parentheses) for the current
experimental upper bound $1/\gamma \ge 2 \, \rm{\mu s}$. Though
the signal-to-noise of the readout is very high in either case,
the estimate of the readout error rate is dominated by the
probability of qubit relaxation during the measurement, which has
a duration of a few cavity lifetimes ($\sim 1-10 \ \kappa^{-1}$).
If the qubit non-radiative decay is low, both high efficiency
readout and more than $10^3$ two-bit operations could be
attained.}
\end{table}
%%%%%%%%%%%%%%%%%%%%%%%%%%%%%%%%%%%%%%%%%%%%%%%%%%%%%%%%%%%%%%%%%%%%%%%%%%%%%%%%%%%%%%%%%%%%%

%\newpage

\end{document}